\documentclass[12pt]{article}

\newcommand {\be}{\begin{equation}}
\newcommand {\ee}{\end{equation}}
\newcommand {\bea}{\begin{eqnarray}}
\newcommand {\eea}{\end{eqnarray}}
\newcommand {\refeq}[1] {(\ref{#1})}

\newcommand {\vett}[1] {\mathbf{#1}}
\usepackage{a4}
\usepackage{psfig, graphicx}
\begin{document}

\date{}
 
\title{VISCO-ELASTIC SPECTRA OF A DILUTE BOSE FLUID}
\author{A. Minguzzi\footnote{Corresponding author. Tel.:+39-050-509058; fax:+39-050-563513; e-mail: minguzzi@ cibs.sns.it}\hspace{0.15cm}  and M. P. Tosi\\
Istituto Nazionale di Fisica della Materia and Classe di Scienze, \\Scuola Normale Superiore, I-56126 Pisa\\ and \\
Abdus Salam International Center for Theoretical Physics,\\ I-34014 Trieste, Italy}

\maketitle
\begin{abstract}
A recently developed local-density current functional formalism for confined Bose-condensed superfluids requires visco-elastic spectra which are defined through a finite-frequency extension of the dissipative coefficients entering the linearized hydrodynamic equations of the two-fluid model. We evaluate these spectra for a superfluid with contact interactions in the collisionless regime at finite temperature, by working to first order beyond the Bogolubov approximation. We find that within this  approximation all the visco-elastic spectra take the same value aside from simple multiplicative factors. 
\end{abstract}
\vspace{4cm}

\newpage
\section{\bf{Introduction}}
Several experiments probing the dynamics of confined Bose-condensed alkali vapours have been reported in the recent literature 
[1 - 6]. 
In this connection, and following the trace set by theoretical work on the dynamics of inhomogeneous electron fluids in the normal state \cite{sette,otto},  a Current-Density Functional  Theory (CDFT) has been developed for an inhomogeneous superfluid in the linear response regime \cite{nove}. In particular, in the case of weak space dependences the CDFT equations take the form of a finite-frequency extension of the linearized hydrodynamics of the two-fluid model, taken at the local equilibrium densities of superfluid and normal fluid. The local-density exchange-correlation kernels entering the CDFT equations are complex functions of frequency describing visco-elastic behaviour, {\it i.e.} each kernel includes both a dissipative spectrum and a frequency-dependent modulus which are related to each other by a Kramers-Kronig relation. These two components yield  in the low-frequency limit  the dissipative friction coefficients of the linearized hydrodynamics of the superfluid and the thermodynamic values of the  velocities of first and second sound, respectively \cite{dieci, undici}.

In further work \cite{dodici} the bulk and shear viscosity spectra of a Bose-condensed fluid have been evaluated by a method previously developed for degenerate electron fluids \cite{tredici, quattordici}. In this approach the equation of motion for the current density is treated by an approximate decoupling to account for the processes of excitation of correlated pairs of particles. One then recovers from the generalized bulk viscosity at resonance  the early results of Beliaev \cite{quindici} and Popov \cite{sedici} for the width of the phonon against decay into two phonons. 

In this  work we present a calculation of all the visco-elastic spectra for a dilute Bose-condensed fluid at finite temperature, following  an approach which was developed in early calculations of the van Hove dynamic structure factor $S(k,\omega)$ by Wong and Gould \cite{diciassette, diciotto}. The interactions between the particles are described by means of a Fermi model potential, {\it i.e.} through a parameter $v=4 \pi a /m$ where $a$ is the $s$-wave scattering length and $m$ is the particle mass ($\hbar=1$). We express the spectra in terms of regular ({\it i.e.} proper and irreducible \cite{diciannove}) functions and then evaluate them by treating the interactions to first order beyond the Bogolubov approximation. As was emphasized by Wong and Gould \cite{diciassette}, by dealing with the regular quantities this approach has two main advantages: (i) the single-particle spectrum and the density fluctuation spectrum coincide, since all the elementary excitations are derived from the zeroes of a suitably defined dielectric function (see also \cite{venti}); and (ii)  particle conservation is ensured by treating directly the longitudinal-current response, to which the density-density and density-current response functions are related {\it via} the continuity equation. Therefore, one is working in an approximation which is both {\it gapless} and {\it conserving}.     

In Section 2 we point out  that within this so-called one-loop approximation the five visco-elastic spectra of the superfluid are determined by the values of four proper ``building blocks''. These are  a condensate kernel related to the shift in local chemical potential due to interactions in the fluid away from equilibrium, two noncondensate kernels describing the regular part of the longitudinal and transverse current-current response, and a cross condensate-noncondensate vertex function. In Section 3 we evaluate these functions and find that the values of the visco-elastic spectra differ only by simple multiplicative factors. We also recover the equation-of-motion results \cite{dodici} for the viscosity spectra. A brief discussion concludes the paper in Section 4. 

\section{Building blocks for the visco-elastic spectra}
We use the standard definition $\chi_{AB}({\vett r-r'},t)=-i \theta(t)\langle [A({\vett r},t),B({\vett r'},0)]\rangle$ for a causal response function relating to two operators in the Heisenberg representation, with $\theta(t)$ the Heaviside step function and $\langle ...\rangle$ the expectation value in the equilibrium ensamble. We also introduce the Fourier transform $\chi_{AB}(k, \omega)$. The visco-elastic spectra of the superfluid are as follows \cite{nove}:
\be
\Re \zeta_3 (\omega)= \lim_{k \rightarrow 0}\frac{- \omega}{k^2} \Im \chi_{{\vett v}_s {\vett v}_s} (k, \omega) \label{equno}
\ee
\be
\Re \zeta_1 (\omega)= \Re \zeta_4 (\omega)=\lim_{k \rightarrow 0}\frac{-m \omega}{k^2} \Im \chi_{\vett {j v}_s}^L (k, \omega)
\ee
\be
\Re \left[\zeta_2 (\omega)+ \frac{4}{3} \eta(\omega) \right]= \lim_{k \rightarrow 0}\frac{- m^2\omega}{k^2} \Im \chi_{\vett {j j }}^L (k, \omega) \label{eqtre} 
\ee
and
\be
\Re \eta(\omega) = \lim_{k \rightarrow 0}\frac{- m^2\omega}{k^2} \Im \chi_{\vett {j j }}^T (k, \omega) \;\;. \label{eqquattro}
\ee
Here, ${\vett v}_s$ is the superfluid velocity and $\vett j$ is the current density, which has been decomposed into its longitudinal ($L$) and transverse ($T$) components. The notation chosen for the spectra serves to remind us that in the limit $\omega \rightarrow 0$ they reduce to the dissipative coefficients entering the linearized hydrodynamic equations of the superfluid (see {\it e.g.} Khalatnikov \cite{dieci}). 

\subsection{Vertex function, regular current response and self-energies}
Wong and Gould \cite{diciassette} studied in detail the longitudinal 
current-current response function in the RHS of Eq. \refeq{eqtre}, which is related to the van Hove structure factor by
\be
\Im \chi_{\vett {j j}}^L(k, \omega) = -\frac{\pi \omega^2}{k^2}S(k, \omega) \;\;.
\ee 
However, they also treated the single-particle Green's function, which has the meaning of an amplitude-amplitude response  for the Bose fluid, and introduced vertex functions relating to it the amplitude-density and amplitude-current response functions. Evidently, the vertex functions connect the condensate with the noncondensate (see also \cite{nove}). As we shall see below, their treatment allows the evaluation of all the spectra in Eqs.~\refeq{equno}-\refeq{eqquattro}.  

The single-particle Green's function is defined by 
\be
G_{\mu \nu}(k, \omega)\equiv \langle \langle a_{{\vett k}\mu}^{};a_{{\vett k}\nu}^{\dag} \rangle \rangle_{\omega}\;\;,
\ee
where $a_{{\vett k}\mu}= a_{\vett k}$ for $\mu=+$, $a_{{\vett k}\mu}= a_{\vett -k}^{\dag}$ for $\mu=-$ and we use the notation 
\be
\langle \langle A;B \rangle \rangle_{\omega}= - i \int_0^{\infty}\, dt \exp(i \omega t) \langle [A(t), B(0)] \rangle \;\;.
\ee
We similarly define the current-amplitude response function 
\be
C_{\mu}^z(k, \omega)\equiv \langle \langle J_{{\vett k}z}^{};a_{{\vett k}\mu}^{\dag} \rangle \rangle_{\omega}
\ee
and the longitudinal and transverse current-current reponse functions 
\be
\chi_{\vett{j j}}^L(k, \omega)=\langle \langle \hat{k}\cdot{ \vett J_k}^{};\hat{k}\cdot{ \vett J_k}^{\dag} \rangle \rangle_{\omega} \label{eqnove}
\ee
and 
\be
\chi_{\vett{j j}}^T(k, \omega)=(\delta_{ij}-\hat{k}_i\hat{k}_j)\langle \langle 
J_{{\vett k}i}^{}; J_{{\vett k}j}^{\dag}\rangle \rangle_{\omega} \;\;. 
\ee
The response functions involving the superfluid velocity are 
\be
\chi_{{\vett v}_s {\vett v}_s} (k, \omega)=(k^2/4m^2 n_0) \beta_{\mu}G_{\mu \nu}(k, \omega)\beta_{\nu} \label{equndici}
\ee
and 
\be
\chi_{\vett{ j v}_s}^L (k, \omega)=(k/2m \sqrt{n_0}) C_{\mu}^z(k, \omega) \beta_{\mu} \label{eqdodici}
\ee
where $\beta_{\mu}= {\rm sgn} (\mu)$ and $n_0$ is the condensate density. Summation over repeated indices is implied in these expressions.

By virtue of the fact that only the long-wavelength limit of the response functions is needed in Eqs.~\refeq{equno}-\refeq{eqquattro}, their evaluation reduces to that of their proper equivalents. An explicit proof 
 is given in Appendix A within the one-loop approximation. We denote  proper quantities by a tilde and regular ({\it i.e.} proper and irreducible) quantities by a $ (r)$ superscript. Following Wong and Gould \cite{diciassette}, we define the vertex function $\tilde \Lambda_{\mu}^z$ by the relation 
\be
\tilde C_{\mu}^z=\tilde \Lambda_{\nu}^z\tilde{G}_{\nu\mu}\;\;. \label{eqtredici}
\ee
The diagrammatic analysis that they give leads to the results
\be
\tilde{\chi}^L_{\vett {j j}}=\tilde{\chi}^{L(r)}_{\vett {j j}}+\tilde \Lambda_{\mu}^z\tilde{G}_{\mu\nu}\tilde \Lambda_{\nu}^z
\ee
and
\be
\tilde{\chi}^T_{\vett{ j j}}=\tilde{\chi}^{T(r)}_{\vett {j j}}\;\;.\label{eqquindici}
\ee
Furthermore, the proper Green's function is expressed in terms of the proper self-energies $\tilde M_{\mu \nu}$ through the Dyson equation as 
\be
\tilde {G}_{\mu \nu}= \tilde {N}_{\mu \nu}/\tilde{D}\;\;,\label{eqsedici}
\ee
where
\be
\tilde N_{\alpha \beta}= \delta_{\alpha \beta}[{\rm sgn}(- \alpha) \omega +\epsilon_k- \mu] + {\rm sgn}( \alpha){\rm sgn}( \beta)M_{-\alpha - \beta}
\ee
and
\be
\tilde D=(\omega -\tilde A)^2- (\epsilon_k- \mu+ \tilde S - \tilde M_2)(\epsilon_k- \mu+ \tilde S + \tilde M_2)\;\;. \label{eqdiciotto}
\ee
In the denominator we have used the symmetrized combinations of the self-energies: $\tilde S= (\tilde M_{++}+\tilde M_{--})/2$, $\tilde A= (\tilde M_{++}-\tilde M_{--})/2$ and  $\tilde M_2= \tilde M_{+-}=\tilde M_{-+}$. Finally, $\epsilon_k = k^2/2m$ and $\mu$ is the chemical potential.

The evaluation of the visco-elastic spectra thus involves the quantities that we have introduced in Eqs.~\refeq{eqtredici}-\refeq{eqsedici}. We proceed to treat these quantities in a  perturbative framework. 

\subsection{One-loop approximation}
In the Bogolubov approximation the excitation spectrum reduces to a single undamped mode at each wavenumber, with a dispersion relation changing continuously from phonons at long wavelengths to free particles at high momenta. The condensate density $n_0$ is equal to the total density $n$ and the total current coincides with the superfluid (or condensate) current. The values taken by the quantities that we have introduced in Section 2.1 are $\tilde S^{(0)}=\mu^{(0)}=vn$, $\tilde M_2^{(0)}=0$, $\tilde A^{(0)}=0$, $\tilde \Lambda_{\mu}^{z(0)}=\sqrt{n_0} k \beta_{\mu}/2m$ and $\tilde \chi^{L,T (r)(0)}=0$.

Broad visco-elastic spectra arise in the so-called one-loop  
approximation, in which one treats 
the corrections to the Bogolubov  results at  first order in the dimensionless coupling-strength parameter $g=((4 \pi a)^3 n)^{1/2}$.
From Eqs.~\refeq{eqtredici}-\refeq{eqdiciotto} we calculate to this order of approximation the proper functions that are needed for the evaluation of the visco-elastic spectra, with the following results:
\bea
\left(\beta_{\mu}\tilde G_{\mu \nu }\beta_{\nu}\right)^{(1)}&=& \frac{2}{\omega^2- \epsilon_k^2} \left[\left(\tilde S^{(1)}+\tilde M_2^{(1)}-\mu^{(1)}\right)\right.\nonumber\\&+& \left. \frac{\epsilon_k}{\omega^2- \epsilon_k^2}\left( 2 \omega \tilde A^{(1)}+ k^2\left(\tilde S^{(1)}-\mu^{(1)} \right)\right) \right]\;\;,\label{eqdiciannove}
\eea
\be
\left(\tilde\Lambda_{\nu}^z\tilde G_{\nu \mu}\beta_{\mu}\right)^{(1)}=
\frac{1}{\omega^2- \epsilon_k^2}\left(\omega \tilde\Lambda_{\mu}^{z(1)} \delta_{\mu}+\epsilon_k \tilde\Lambda_{\mu}^{z(1)} \beta_{\mu}\right)+ \frac{k}{2}\left(\beta_{\mu}\tilde G_{\mu \nu }\beta{\nu} \right)^{(1)}\;\;,
\ee
\bea
\tilde \chi_{\vett{ jj}}^{L(1)}&=&
\tilde \chi_{\vett {jj}}^{L(r)(1)}+
\frac{\epsilon_k}{\omega^2- \epsilon_k^2} 
\left[-\epsilon_k n'^{(1)}+ \frac{2 \omega}{k} \tilde \Lambda_{\mu}^{z(1)} 
\delta_{\mu} + k \tilde \Lambda_{\mu}^{z(1)} \beta_{\mu}
\right. \nonumber \\  
&+& \left.\left(\tilde S^{(1)}+\tilde M_2^{(1)}-\mu^{(1)} \right)+ 
\frac{\epsilon_k}{\omega^2- \epsilon_k^2}
\left(2 \omega \tilde A^{(1)}+ k^2 (\tilde S^{(1)}-\mu^{(1)})\right) 
\right] \label{eqventuno}
\eea
and
\be
\tilde \chi_{\vett{ jj}}^{T(1)}=
\tilde \chi_{\vett {jj}}^{T(r)(1)}\;\;. \label{eqventidue}
\ee
In these equations $\delta_{\mu}$ is the vector $(1,1)$ and $n'=n-n_0$ is the depletion of the condensate. We have used the convention of summation over repeated indices and expressed wavenumbers in units of $\sqrt{4 \pi n a}$ and energies in units of $4 \pi n a/m$. In these units the coupling strength parameter is $g=4 \pi a=1/n$. 

It is evident that the evaluation of  Eqs.~\refeq{eqdiciannove}-\refeq{eqventidue} involves seven quantities, that is $(\tilde S^{(1)}+\tilde M_2^{(1)}-\mu^{(1)})$, $A^{(1)}$, $(\tilde S^{(1)}-\mu^{(1)})$, $\tilde\Lambda_{\mu}^{z(1)} \delta_{\mu}$, $\tilde\Lambda_{\mu}^{z(1)} \beta_{\mu}$, $\tilde \chi_{\vett {jj}}^{L(r)(1)}$ and $\tilde \chi_{\vett {jj}}^{T(r)(1)}$.
The relevant one-loop diagrams for these  functions are shown in Figure 1, where the single-particle propagators represent Bogolubov quasiparticles. In the present units each condensate line carries a factor $g^{-1/2}$ and each current vertex carries a factor $g^{1/2}$. The self-energy diagrams in Figure 1 include two-quasiparticle excitation processes {\it via} scattering against the condensate, in addition to the usual Hartree and Fock terms as for a normal fluid.   
The vertex function accounts for damping of longitudinal-current fluctuations {\it via} excitation of quasiparticles and scattering against the condensate. Finally, the attenuation of both longitudinal and transverse current fluctuations  proceeds {\it via} two-quasiparticle processes. 

We show in Appendix B that the quantities  $A^{(1)}$, $(\tilde S^{(1)}-\mu^{(1)})$ and  $\tilde\Lambda_{\mu}^{z(1)} \beta_{\mu}$ do not contribute to the visco-elastic spectra in the long wavelength limit. Therefore, only the four quantities $(\tilde S^{(1)}+\tilde M_2^{(1)}-\mu^{(1)})$, $\tilde\Lambda_{\mu}^{z(1)} \delta_{\mu}$, $\tilde \chi_{\vett {jj}}^{L(r)(1)}$ and $\tilde \chi_{\vett {jj}}^{T(r)(1)}$ remain to be calculated in that limit. 
The first of these is a condensate kernel contributing to  all three longitudinal spectra, while the second is a cross condensate-noncondensate kernel involved in the longitudinal current spectrum and in the correlations between longitudinal current and superfluid velocity fluctuations. Finally, two noncondensate kernels enter the  attenuation of  longitudinal and transverse current fluctuations.

\section{Calculation of visco-elastic spectra}
The calculation of the diagrams shown in Figure 1 yields the following expressions for the building blocks of the visco-elastic spectra at zero temperature  \cite{diciassette,diciotto}:
\be
\Im(\tilde S^{(1)}+\tilde M_2^{(1)}-\mu^{(1)})=- \pi  \int \, \frac{d^3 p}{(2 \pi)^3 }\lambda_{\vett p} \lambda_{\vett{ p+k}} \delta (\omega-\omega^0_{\vett{p+k}}-\omega^0_{\vett{p}})\;,\label{eqventitre}
\ee
\be
\Im(\tilde\Lambda_{\mu}^{z(1)} \delta_{\mu})=-\frac{\pi}{2}\int \, \frac{d^3 p}{(2 \pi)^3 }\left({\vett p}\cdot \hat k + \frac{1}{2}k\right) \left(\lambda_{\vett p}- \lambda_{\vett{ p+k}} \right) \delta (\omega-\omega^0_{\vett{p+k}}-\omega^0_{\vett{p}})\;,
\ee
\be
\Im(\tilde \chi_{\vett {jj}}^{L(r)(1)})= -\frac{\pi}{8}\int \, \frac{d^3 p}{(2 \pi)^3 }\left({\vett p}\cdot \hat k + \frac{1}{2}k\right)^2 \frac{(\lambda_{\vett p}- \lambda_{\vett{ p+k}} )^2}{\lambda_{\vett p} \lambda_{\vett{ p+k}}} \delta (\omega-\omega^0_{\vett{p+k}}-\omega^0_{\vett{p}})
\ee
and
\be
\Im(\tilde \chi_{\vett {jj}}^{T(r)(1)})= -\frac{\pi}{16}\int \, \frac{d^3 p}{(2 \pi)^3 }\left[ p^2 -({\vett{p}} \cdot \hat k )^2\right] \frac{(\lambda_{\vett p}- \lambda_{\vett{ p+k}} )^2}{\lambda_{\vett p} \lambda_{\vett{ p+k}}} \delta (\omega-\omega^0_{\vett{p+k}}-\omega^0_{\vett{p}})\;.\label{eqventisei}
\ee
In these equations we have defined $\omega^0_{\vett{p}}=\sqrt{p^2+ p^4/4
}$ and $\lambda_{\vett p}=\epsilon_{\vett p}/\omega^0_{\vett{p}}$, $\omega^0_{\vett{p}}$ being the Bogolubov quasiparticle frequency in reduced units. 

The thermal  factor $(1+f_{\vett {p+k}}+f_{\vett {p}})$, with  $f_{\vett {p}}=(\exp(\beta \omega^0_{\vett{p}})-1)^{-1}$  and $\beta=1/k_BT$, must be inserted in the integrands in Eqs.~\refeq{eqventitre}-\refeq{eqventisei} at finite temperature, on account of detailed balance in the process of decay of an excitation of energy $\omega$ into two  quasiparticles. In addition, there appear  other contributions  associated with the scattering of the excitation against a thermally excited quasiparticle. However, the corresponding integrals contain a factor $(f_{\vett {p+k}}-f_{\vett {p}})$, which vanishes in the long wavelength limit. 

Bearing in mind the above observations, we easily find from Eqs.~\refeq{eqventitre}-\refeq{eqventisei} the following expressions for the four building blocks in the long wavelength limit at finite temperature:
\be
\lim_{k \rightarrow 0}\Im(\tilde S^{(1)}+\tilde M_2^{(1)}-\mu^{(1)})=\left. -\frac{\sinh(\beta \omega/2)}{\cosh(\beta \omega/2)-1}\;\frac{p^4}{16 \pi\omega_{p}' (1+p^2/4) }\right|_{\omega^0_p=\omega/2}\;,
\ee
\be
\lim_{k \rightarrow 0}\Im(\tilde\Lambda_{\mu}^{z(1)} \delta_{\mu})=\left.\frac{\sinh(\beta \omega/2)}{\cosh(\beta \omega/2)-1}\; \frac{k p^3}{48 \pi\omega_{p}' (1+p^2/4)^{3/2} }\right|_{\omega^0_p=\omega/2}
\ee
and
\bea
\lim_{k \rightarrow 0}\Im(\tilde \chi_{\vett {jj}}^{L(r)(1)})&=&
\left. -\frac{\sinh(\beta \omega/2)}{\cosh(\beta \omega/2)-1}\;\frac{k^2 p^2}{160  \pi\omega_{p}' (1+p^2/4)^2 }\right|_{\omega^0_p=\omega/2}\nonumber \\ &=& 3 \lim_{k \rightarrow 0}\Im(\tilde \chi_{\vett {jj}}^{T(r)(1)})\;. 
\eea
Here, $\omega_{p}'=d\omega_p^0/dp$ and  $p$ is defined by the condition $\omega^0_p=\omega/2$.

Using these results in Eqs.~\refeq{eqdiciannove}-\refeq{eqventidue} and hence in Eqs.~\refeq{equndici}-\refeq{eqquindici}, we obtain 
\bea
\Re\left(\eta(\omega)\right)&= &\frac{4}{7} \Re\left(\zeta_2(\omega)+\frac{4}{3}\eta(\omega)\right)= \frac{4}{5}nm\Re\left(\zeta_1(\omega) \right)= \frac{4}{15}n^2 m^2\Re\left(\zeta_3(\omega)\right)\nonumber  \\ &=& \left.\frac{\sinh(\beta \omega/2)}{\cosh(\beta \omega/2)-1}\;\frac{n^2v^2 p^6} {30 \pi m^2 {\omega}^3 \omega_p'}\right|_{\omega^0_p=\omega/2}\;\;.\label{eqtrenta}
\eea
We remark that the behaviour of these spectra at low frequency is proportional to $\omega^3$  at $T=0$ and  to $\omega^2 T$ at finite temperature. As already noticed in the Introduction, the result for the shear and bulk viscosity spectra agrees with that  obtained in \cite{dodici} by an equation-of-motion approach in an approximation which takes account of  the excitation processes of two quasiparticles.   

Figure 2 shows the behaviour of the shear viscosity spectrum at zero temperature and at finite values of the temperature. We remark from Eq.~\refeq{eqtrenta} that the spectra involve the Fourier transform $v$ of the interatomic potential, so that their high-frequency behaviour depends on this potential at high momenta and hence  on the details of the short-range interatomic collisions.  Difficulties can be anticipated in assessing this asymptotic behaviour for the Bose-condensed alkali vapours: in this case a realistic description of the short-range potential (a) is not Fourier transformable, and (b) gives a bound state which is not to be included in the study of the metastable condensate state. 

A simple way to regularize the contact potential that we have assumed in the present calculations would be through the introduction of a high-momentum cut-off \cite{dodici}. For a reasonable value of the cut-off the results shown over the frequency range illustrated in Figure 2, as well as the simple proportionality of the various spectra which is seen in Eq.~\refeq{eqtrenta}, would remain approximately valid. 

The Hilbert transform of the visco-elastic spectra gives the frequency-dependent corrections to the thermodynamic values of the moduli determining the first and second sound velocity. Evidently, regularization of the interaction potential will be necessary before evaluating these corrections. 

\section{Concluding remarks}
We have calculated the visco-elastic spectra for a dilute Bose-condensed fluid in the collisionless regime at finite temperature. The motivation for this work has been provided by the theory developed in \cite{nove}, showing that these spectra as determined  for  a homogeneous Bose fluid allow a local density treatment of  the dynamics of a weakly inhomogeneous superfluid. 

We have assumed contact  interactions between the particles and treated  them to first order beyond the Bogolubov approximation. Within this scheme   we have found that in the long-wavelength limit  the spectra  are determined by four proper  building blocks, to which we have attributed meaning in terms of dynamic correlations between condensate and noncondensate. Explicit evaluation of these four functions has shown that within this approach the visco-elastic spectra are in fact determined by a single function of frequency. However, the high-frequency part of the spectra and the  visco-elastic shifts in the sound velocity moduli will be sensitive to details of short-range interatomic collisions beyond the contact-potential scheme.  

We have briefly indicated how our calculation of the  shear and bulk viscosity spectra  for the superfluid is related to an equation-of-motion treatment of the dynamics of current fluctuations. The latter approach has also been applied to an electron  fluid in the normal state. Since  we have treated the dissipation spectra in the superfluid as arising from the interactions between  Bogolubov quasiparticles, we have found that the current-density fluctuations  can only decay into pairs of longitudinal excitations.   In contrast,  decay channels invoving transverse current fluctuations also  contribute to damping in a   normal fluid. 

 Experimental studies of higher frequency excitations in the collisionless regime would be useful  for quantitative tests of the theory. 
 The presently available experiments on confined Bose-Einstein condensates refer to low-frequency excitations, in  a dynamical regime  where collisions begin to become important depending on the thermodynamic state of the fluid. 
The theory of the transition to a  collisional  regime, where the dynamical behaviour of the fluid is governed by a Boltzmann equation at frequencies  below an average collision rate, remains to be explored. A start in this direction was made in the early work of Popov \cite{sedici}.

\subsection*{Acknowledgements}

This work is supported by the Istituto Nazionale di Fisica della Materia through the Advanced Research Project on BEC.

\appendix{}
\section*{Appendix A. Reduction to  proper quantities}
In this Appendix we demonstrate that within the one-loop approximation the response functions introduced in Eqs.~\refeq{eqnove}-\refeq{eqdodici} reduce to the corresponding proper response functions in the long wavelength limit.

We start from the superfluid velocity response in Eq.~\refeq{equndici}. From the diagrammatic analysis given by Wong and Gould  \cite{diciassette} for the single-particle Green's function one easily finds 
\be
\beta_{\mu} G_{\mu \nu} \beta_{\nu}= \frac{\epsilon^{(r)}}{\epsilon}\beta_{\mu} \tilde G_{\mu \nu} \beta_{\nu} +\frac{v}{\tilde D \epsilon}\left(\frac{k}{\omega}\tilde \Lambda_{\mu}^z \delta_{\mu}+ \frac{2 \sqrt{n_0}}{\omega}(\omega-\tilde A)\right)^2
\ee
where $\epsilon$ has the meaning of a dielectric function in terms of the proper density-density response function and $\epsilon^{(r)}$ is its regular part. A lengthy but straightforward calculation yields the one-loop result
\bea
\Im (\beta_{\mu} G_{\mu \nu} \beta_{\nu})^{(1)}&=& \frac{\omega^2-\epsilon_k^2}{\omega^2-{\omega_k^{0}}^2}\Im  (\beta_{\mu}\tilde G_{\mu \nu} \beta_{\nu})^{(1)}
+\frac{4k}{\omega(\omega^2-{\omega_k^{0}}^2) }\tilde \Lambda_{\mu}^{z(1)} \delta_{\mu}\nonumber \\&+& \frac{2 \epsilon_k}{\omega^2 (\omega^2-{\omega_k^{0}}^2)}\Im\left[ 2 k \tilde\Lambda_{\mu}^{z(1)}\beta_{\mu}-2 (\tilde S^{(1)} + \tilde M_2^{(1)} -\mu^{(1)}) \right.\nonumber\\ &+&\left.4 (\omega^2-\epsilon_k^2)\tilde \chi_{\vett{jj}}^{L(1)}-k^2 \tilde \chi_{\vett{jj}}^{L(r)(1)}+\frac{4 \omega^2}{\omega^2-\epsilon_k^2}(\tilde S^{(1)} -\mu^{(1)})\right]\nonumber\\&+&\frac{4 \epsilon_k^2}{\omega^2(\omega^2-{\omega_k^{0}}^2)}\Im \left[(\omega^2-\epsilon_k^2)\tilde \chi_{\vett{jj}}^{L(1)}+\frac{2 \omega}{\omega^2-\epsilon_k^2}\tilde A^{(1)} \right]
\;,\label{a2}
\eea
in terms of the functions  entering Eqs.~\refeq{eqdiciannove}-\refeq{eqventuno}. 
The long wavelength behaviour of these functions, from Eqs.~\refeq{eqventitre}-\refeq{eqventisei} and from the results given in Appendix B, is as follows:
all the self-energy terms are of order $k^0$, the vertex-function terms are of order $k^1$ and the current response functions are of order $k^2$. Hence, from Eq.~\refeq{a2} it follows that 
\be
\lim_{k \rightarrow 0}\Im (\beta_{\mu} G_{\mu \nu} \beta_{\nu})^{(1)}= \lim_{k \rightarrow 0}\Im  (\beta_{\mu}\tilde G_{\mu \nu} \beta_{\nu})^{(1)}
\ee

Similarly for the ${\vett{ J-v}}_s$ response function we have 
\bea
\frac{k}{2}\Im\left(C_{\mu}^z\beta_{\mu}\right)^{(1)}&=& \frac{k}{2}\Im\left(\tilde\Lambda_{\nu}^z\tilde G_{\nu \mu}\beta_{\mu}\right)^{(1)}+\frac{\epsilon_k^2}{\omega^2(\omega^2-\epsilon_k^2)}\Im \left[k^2 \tilde \chi_{\vett{jj}}^{L(r)(1)}\right.\nonumber \\ &-& \left.2 k \tilde \Lambda_{\mu}^{z(1)}\beta_{\mu}+2 (\tilde S^{(1)} + \tilde M_2^{(1)} -\mu^{(1)})\right]
\eea
and one immediately finds 
\be
\lim_{k\rightarrow 0}\Im (C_{\mu}^z \beta_{\mu})^{(1)}=\lim_{k\rightarrow 0}\Im( \tilde C_{\mu}^z \beta_{\mu})^{(1)} \;.
\ee

Finally, for the current response function one can use directly Eqs.~(3.31) and (3.32) in the paper of Wong and Gould \cite{diciassette} to obtain
\be
\lim_{k\rightarrow 0}\Im \chi_{\vett{jj}}^{L(1)} = \lim_{k\rightarrow 0}\Im\tilde \chi_{\vett{jj}}^{L(1)}\;.
\ee
A similar property is evidently valid for $\chi_{\vett{jj}}^{T}$.

\section*{Appendix B. Some further one-loop results}
We complete the calculation of the one-loop spectral functions in supplement to Eqs.~\refeq{eqventitre}-\refeq{eqventisei}. We have
\be
\Im \tilde A^{(1)}=-\pi \int \, \frac{d^3 p}{(2 \pi)^3 } \lambda_{\vett p} \delta (\omega-\omega^0_{\vett{p+k}}-\omega^0_{\vett{p}})\;, \label{b1}
\ee
\be
\Im (\tilde S-\mu)^{(1)}=- \frac{\pi}{2}\int \, \frac{d^3 p}{(2 \pi)^3 } \left[\lambda_{\vett p}\lambda_{\vett{ p+k}}+\frac{1}{2}\left(1+ \frac{\lambda_{\vett p}}{\lambda_{\vett{ p+k}}}\right) \right) \delta (\omega-\omega^0_{\vett{p+k}}-\omega^0_{\vett{p}})\label{b2}
\ee
and
\be
\Im( \tilde \Lambda_{\mu}^z \beta_{\mu})^{(1)}= -\frac{\pi}{2}\int \, \frac{d^3 p}{(2 \pi)^3 }\left({\vett p}\cdot \hat k + \frac{1}{2}k\right) \left(\frac{\lambda_{\vett p}}{ \lambda_{\vett{ p+k}}}-1 \right) \delta (\omega-\omega^0_{\vett{p+k}}-\omega^0_{\vett{p}})\;. \label{b3}
\ee

From these expressions it is easily seen that in the long wavelength limit  the quantities \refeq{b1} and \refeq{b2} are of order  $k^0$ and the quantity \refeq{b3} is of order $k^1$. Therefore, their contributions to Eqs.~\refeq{eqdiciannove}-\refeq{eqventuno} vanish in this limit.


\begin{thebibliography}{10}
\bibitem{diciannove}
We recall that a proper diagram cannot be broken into separate pieces by
  cutting an interaction line, while an irreducible diagram cannot be broken
  into separate pieces by cutting a propagator line.
\end{thebibliography}

\newpage
\subsection*{Figure captions}
\vspace{1cm}

\hspace{0.5cm}  {\bf Figure 1}. One-loop diagrams for the proper self-energies $\tilde M_{\mu \nu}$,  vertex function $\tilde \Lambda_{\mu}^z$ and current-current response functions $\tilde \chi_{\vett j j}$. The full lines  are Bogolubov single-particle propagators, while the  wiggly lines represent the condensate and the triangles denote  current vertices. The interaction lines are shown in dashes.
\vspace{1cm}

{\bf Figure 2}. Viscosity spectrum $\eta(\omega)$ at zero (solid line) and finite temperature (dashed $T=3 \omega_0/8$, dot-dashed $T=3 \omega_0/4$) as a function of frequency in units of $\omega_0= 4\pi a n/m$.

\newpage
\begin{figure}
\begin{center}
\includegraphics[scale=0.58]{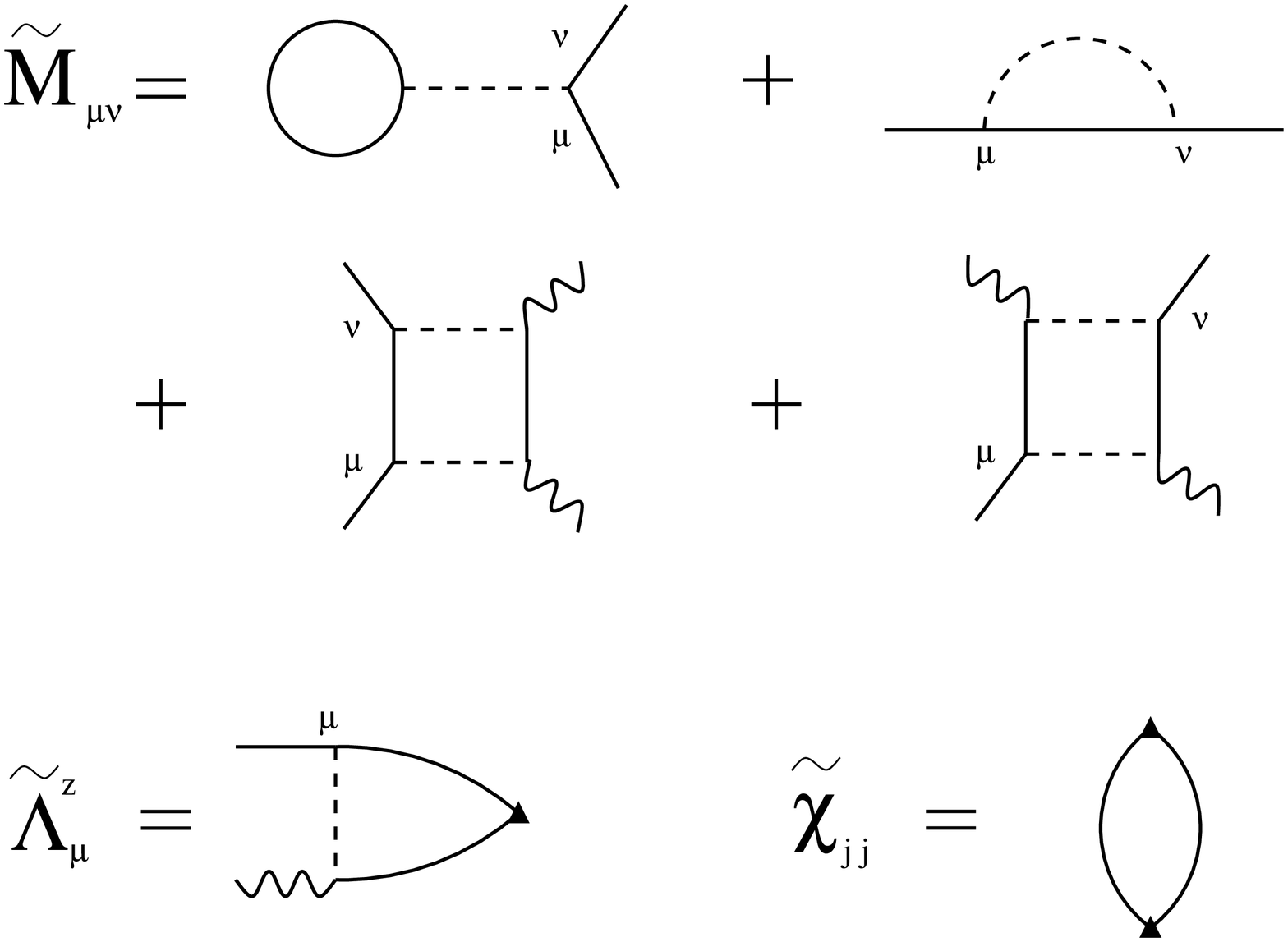} 
\end{center}
\end{figure}
\newpage
\begin{figure}
\begin{center}
\includegraphics[scale=0.7]{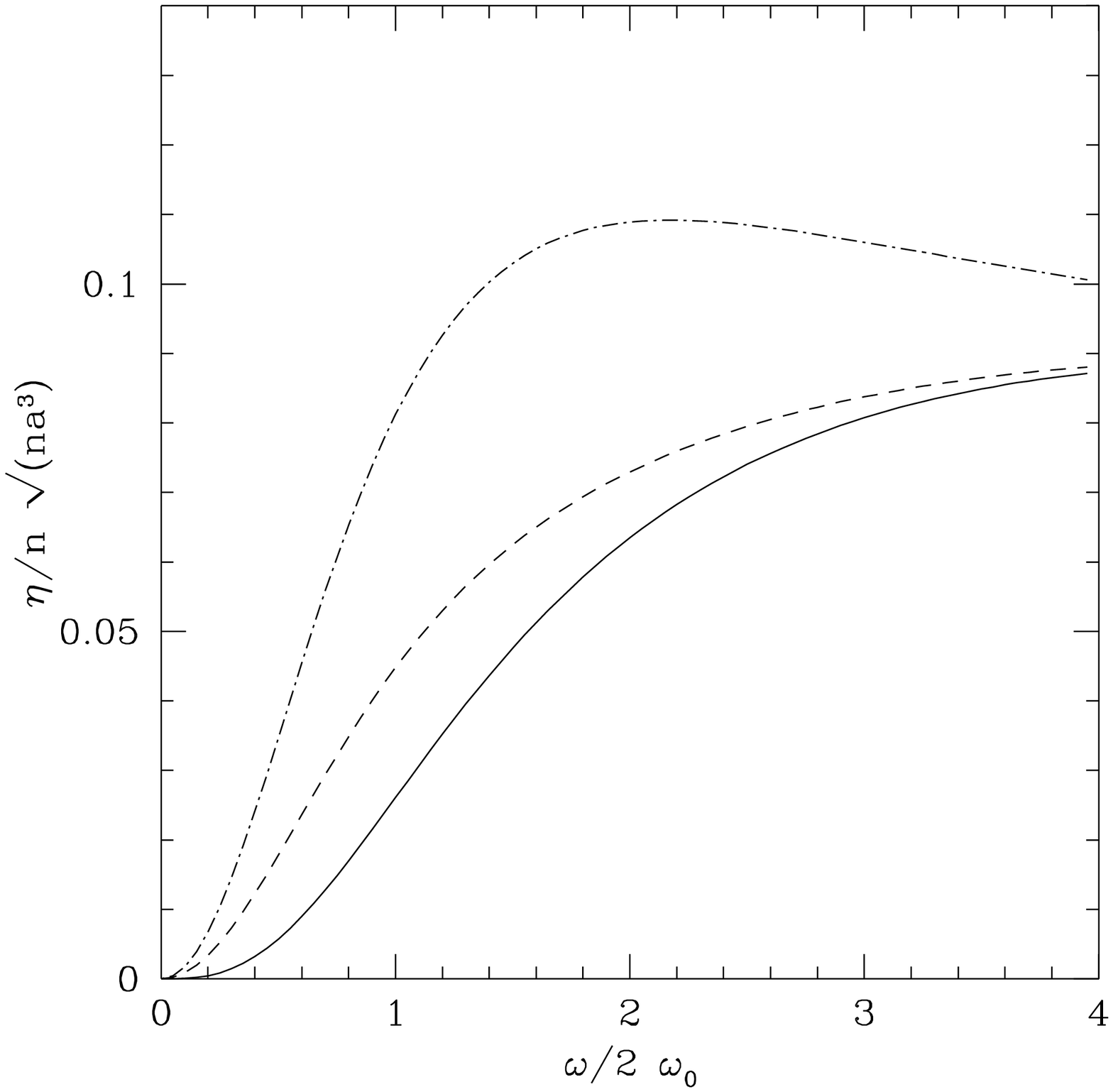} 
\end{center}
\end{figure}

\end{document}